\documentclass{xray_symp}
\usepackage{graphics,psfig}

\begin{document}
\title{AGN/galaxy separation in the ROSAT Bright Survey}
\author{I. Lehmann\inst{1} \and G. Hasinger\inst{1} \and A. D. Schwope\inst{1}
       \and Th. Boller\inst{2}}
\institute{Astrophysikalisches Institut Potsdam,
       An der Sternwarte 16, 14482 Potsdam, Germany
       \and  Max--Planck--Institut f\"ur extraterrestrische Physik,
       Giessenbachstra{\ss}e, 85740 Garching, Germany}

\maketitle

\begin{abstract}
The X-ray luminosity function (XLF) of
galaxies is dominated by AGN (classified by their optical spectra) above
L$_{\mbox{x}}=10^{42}\,\mbox{erg}\,\mbox{s }^{-1}$, below this value by normal 
galaxies. The X-ray flux of AGN at low X-ray luminosity therefore contains contributions from stellar processes (winds, supernova explosions etc.). Until now it is not clear which fraction of the observed X-ray flux has to be associated to the underlying stellar emission.
We started to investigate a complete luminosity-limited sample of AGN and galaxies with log L$_{\mbox{x}}\le42.5$ to discriminate between the nuclear and
galaxy flux. Using the ROSAT HRI data we calculated the radial
wobble-corrected profiles of the sources which were compared to a new HRI point
spread function (PSF) template derived from 21 X-ray bright stars.
The majority of the low-luminosiy AGN shows evidence for a
significant fraction of extended X-ray emission.
\end{abstract}

\section{Introduction}

The soft X-ray background (XRB) is dominated by the integrated emission
of discrete sources (Hasinger et al. 1998, Hasinger 1998). Active galactic
nuclei (AGN) are dominating the source counts down to the faintest flux
limits (Schmidt et al. 1998). However, there are indications, that X-ray active, optically normal galaxies start to dominate the X-ray background at the faintest fluxes (McHardy et al. 1998), which may be obscured AGN.

\begin{figure}[hb]
\resizebox{0.90\hsize}{!}{\includegraphics{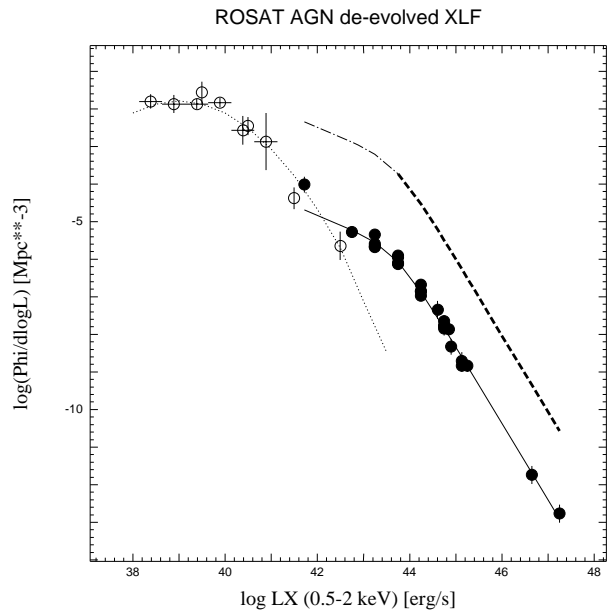}}
 \caption[]{The new XLF of AGN (solid circles) and normal galaxies
   (open circles). The solid line gives the AGN luminosity function de-evolved
   with a $(1+z)^{5}$ density evolution law. The dashed line shows the
   same function at z$=$1.6, where the thick dashed portion indicates the
   observed luminosity interval. The dot-dashed line is an extrapolation, which is 
   currently still uncertain. The XLF of AGN and galaxies fit well together in the
   overlapping region.}
 \label{fig:XLF}
\end{figure}

An X-ray luminosity function has first been determined for AGN, combining
the ROSAT Bright Survey (RBS) and the ROSAT Deep Survey (Hasinger et al. 1998) samples with the
RIXOS AGN sample (Page et al. 1996) and the Appenzeller AGN Sample (Appenzeller
et al. 1998). There is a clear evolution in the AGN luminosity function in the
sense that higher redshift AGN are much more abundant or luminous than the
local sample. In contrast to Boyle et al. (1994) we cannot fit our data
with a pure luminosity evolution, but find a better fit with a density evolution.
There is some indication of a slower evolution at luminosities below 10$^{44}$ 
erg s$^{-1}$ (cf. Schmidt et al. and Miyaji, Hasinger \& Schmidt, same issue) . However, this takes not into account the known soft X-ray absorption of
AGN, which complicates the self-consistent derivation of an XLF model.
Additional uncertainties exist at the low-luminosity end of the local AGN
luminosity function, where X-rays of non-AGN origin start to play an important
 role. 
We have derived an X-ray luminosity function for non-AGN galaxies 
(hereafter ``galaxies``) in the RBS.
The data are shown in Fig. \ref{fig:XLF} with open circles with horizontal error bars.
In addition to our flux-limited sample the local galaxy XLF has been derived
from a volume-limited sample of galaxies within 7.2 Mpc (Boller, Schmidt \& Voges 1998) shown by open circles with horizontal and vertical error bars in Fig. \ref{fig:XLF}.

The two galaxy XLFs fit very nicely together in the overlapping region. From Fig. 
\ref{fig:XLF} it becomes obvious that the total XLF is dominated by AGN above and by galaxies
below an X-ray luminosity of log L$_{X}=$~42.5. The local volume density of AGN is comparable to that of normal galaxies around an X-ray luminosity of 
log L$_{X}=$~42.5.
The question therefore arises, wether the soft X-ray luminosity in these
objects is contaminated by X-ray photons from the host galaxies (visa versa for galaxies).

\section{The ROSAT HRI observations}

We have started to analyse HRI data of all AGN and galaxies in the ROSAT Bright
Survey (Fischer et al. 1998, Schwope et al., in prep)
which have X-ray luminosities below log L$_{X}<$ 42.5 (erg/s). The cassification of the objects is deduced from optical spectroscopy. Tab. \ref{tab:RBS_sa} summarizes the information for the statistically complete luminosity-limited sample of 33 RBS objects, including some well-known AGN
and galaxies eg. NGC 4051, NGC 4151 and M82. 
For 27 RBS sources sufficient HRI data were available from the public ROSAT archive to compare their radial profiles with the ROSAT HRI point-spread-function (PSF).

Because the HRI PSF is significantly contaminated by systematic attitude residuals introduced by the spacecraft wobble we developed a new method to correct for these (Harris et al. 1998). The method is based on the following steps:
\begin{itemize}
     \item Extraction of the HRI data within a constant roll angle interval
       (using the same guide star configuration)
     \item Folding the data over the ROSAT 402 sec wobble phase
     \item Determination of the centroid in phase resolved images
     \item Shifting the phase resolved images to a single centroid position 
       (shift and add technique)
\end{itemize}

We checked this new wobble-correction method on a large number of HRI observations of stars, and it works sucessfully down to HRI countrates of about 0.1 cts/s. 
The wobble-corrected stellar profiles of 21 RBS stars were used to derive a 
re-calibrated HRI PSF template, which is available upon request from the authors.

\begin{figure}[ht]
\resizebox{0.95\hsize}{!}{\includegraphics{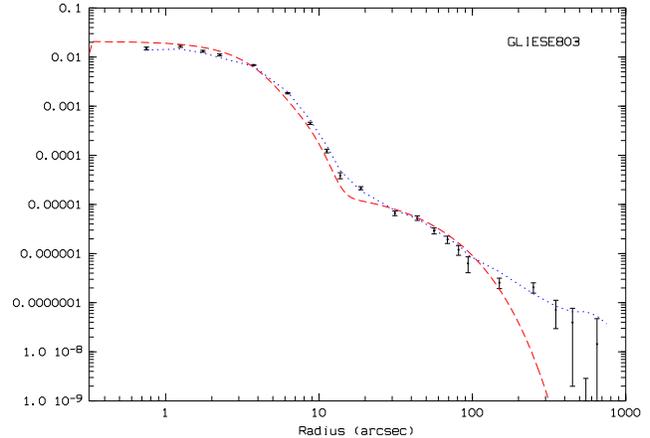}}
\caption[]{Wobble-corrected radial profile of the RBS star GLIESE 803                     normalized to an integrated flux of one
             in comparison to the theoretical HRI PSF (dashed line) and
             the re-calibrated HRI PSF (dotted line).}
\label{fig:prof}
\end{figure}
 Fig. \ref{fig:prof} shows the wobble-corrected
radial profile of the RBS star GLIESE 803 in comparison to the theoretical HRI PSF from the ROSAT User Handbook and to the re-calibrated HRI PSF. The new PSF template fits the data points well. Both HRI PSF differ significantly in the range between 10\arcsec~and 30 \arcsec. 
The deviation at radii larger then 100\arcsec~is due to the ROSAT mirror scattering term.
 To determine a real source extension for slightly extended X-ray sources we recommend to use the more realistic re-calibrated HRI PSF. 

We have derived the radial profiles of 27 nearby low-luminosity AGN and galaxies.
The radial profiles were normalized to an
integrated flux equal one. If an X-ray source would differ from a point source, then the innermost data points of the radial profile would be located below the HRI
 PSF, while the outer data points (r$>$6\arcsec) lie above the curve of the PSF.

\begin{figure}[htp]
\resizebox{0.95\hsize}{!}{\includegraphics{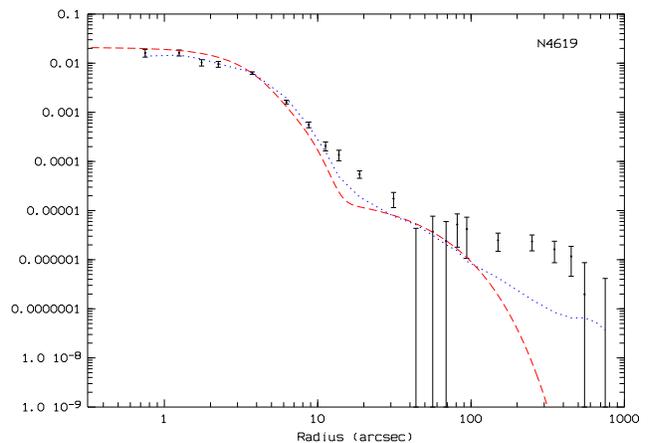}}
\caption[]{The radial profile of NGC 4619 (Sy1) compared to the theoretical HRI PSF and the re-calibrated HRI PSF.
}
\label{fig:rad_sec}
\end{figure}

\begin{figure}[ht]
\resizebox{0.95\hsize}{!}{\includegraphics{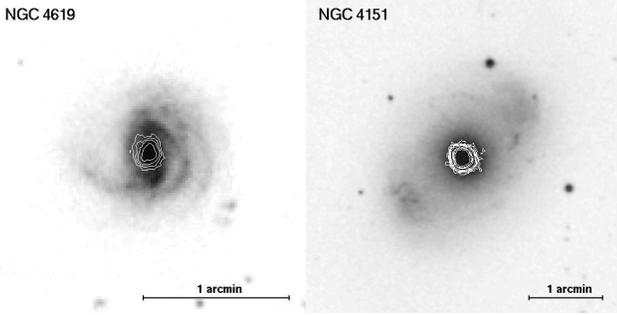}}
\caption[]{HRI contour lines of the AGN NGC 4619 and NGC 4151 superposed on copies of digital Schmidt plates (POSSII) showing the X-ray extended nuclear region.}
\label{fig:cont}
\end{figure}

Fig. \ref{fig:rad_sec} presents the radial wobble-corrected 
profile of the Seyfert 1 galaxy NGC 4619 in comparison to the
HRI PSF's. The radial profile indicates an extented soft X-ray emission up to a radius of about 30\arcsec. The excessive X-ray flux at r$>$100\arcsec~is probably due to the uncertainty in background subtraction. The HRI contours of NGC
4619 show a clearly elongated X-ray emission in north-south direction (cf. Fig \ref{fig:cont}), while the Seyfert 1 galaxy NGC 4151 seems to be only slightly extended in north-east direction (north is on top, east at left). 

Elvis et al. (1990) mentioned that such an extended X-ray emission is presumable a common feature in
Seyfert galaxies. 
Using Einstein HRI data they discovered a fraction of 15-30 \% extended emission in three (NGC 4151, NGC 1599, NGC 2992) out of five Seyfert galaxies. There are various possible orgins for this emission, eg. thermal bremsstrahlung, synchrotron
radiation, inverse Compton scattering and electron-scattered nuclear radiation.  
Wilson et al. (1992) favour for the famous Seyfert galaxy NGC 1068 a hot (10$^{6-7}$ K), outflowing wind as the source of the circumnuclear soft X-ray emission. 

\begin{figure}[htp]
\resizebox{0.90\hsize}{!}{\includegraphics{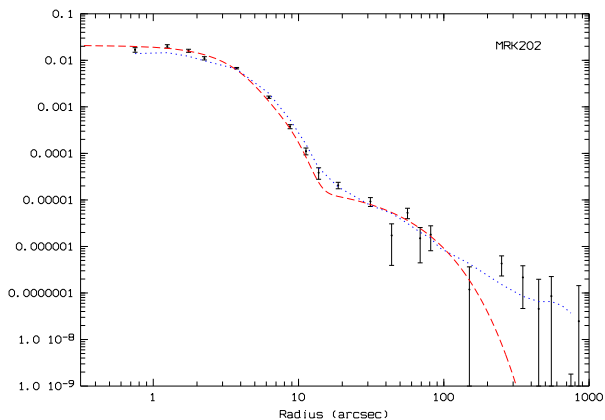}}
\caption[]{The radial profile of the starburst galaxy MRK 202 indicates no
           extended X-ray emission. The data points are consistent with the 
           re-calibrated HRI PSF.}
\label{fig:rad_gal}
\end{figure}

Most of our studied low-luminosity AGN show an indication of an extended
X-ray emission in their radial profiles. Clearly, most of the galaxies reveals 
extended emission in their the radial profiles. On the other hand, we found three examples for point-like X-ray sources in otherwise normal 
galaxies (cf. Fig. \ref{fig:rad_gal}).  
Iyomoto et al. (1998) presented the ASCA results on the S0 galaxy NGC 4203, where they detect hard X-ray emission from a point-like source at the nucleus.
A single power-law model with a photon index of $\approx$1.8, which is a typical
value for Seyfert galaxies, fits the spectrum of NGC 4203 well. A point-like soft X-ray emission of the source was already found by Bregman et al. (1995) using
ROSAT HRI data. 

The results on NGC 4203 strongly suggest that this object hosts a low-luminosity
AGN. In addition to MRK 202, the example shown in Fig. \ref{fig:rad_gal}, the point-like X-ray emission of NGC 1313 seems to indicate an optically hidden AGN (cf Tab. \ref{tab:RBS_sa}). Recently, Ptak et al. (1998) have investigated a low-luminosity sample of AGN and starburst galaxies, including NGC 253, NGC 3147, NGC 3998 and M82, which belong to our sample. For all sources they found evidence in the ASCA hard X-ray spectra for a hard component, which can be explained either with an absorbed power-law or with thermal bremsstrahlung of T$\approx$ 10$^{8}$ K. Future hard X-ray spectroscopy of MRK 202 and NGC 1313 is neccessary to confirm the AGN nature of these objects. 

\section{AGN/galaxy separation in X-rays}

In order to estimate the fraction of extended soft X-ray emission for each object of our sample we used the simple method by Wilson et al. (1992). We subtracted
a suitably scaled re-calibrated HRI PSF from the observed radial profiles. The
residuals of the X-ray flux divided by the total flux within a radius of 100\arcsec~is regarded as the fraction of extended emission. An outer radius of 100\arcsec~was chosen to avoid additional errors due to uncertainties in the background subtraction.
The fraction of extended emission is given in the last column of Tab. \ref{tab:RBS_sa} for the statistically complete RBS AGN/galaxy sample. The classification of
the objects in AGN (ID$=$1) and galaxies (ID$=$2) was done by their optical spectra. The CR and CH values are the total and hard band (0.5-2.0 keV) ROSAT PSPC countrates. The X-ray luminosity is based on the CH countrates.  
Sources without sufficient ROSAT HRI data are scheduled with the HRI in ROSAT AO8. 

The fraction of extended X-ray flux due to our analysis is in good agreement with 
those found in the literature. Elvis et al. (1990) derived for NGC 1566 and NGC 4151 a fraction of 15-30 \% using HRI Einstein data, Wilson et al. (1992) estimated 
for NGC 1068 a value of 45$\pm$5~\%. Therefore we believe that our
X-ray analysis leads to reliable results.

About half of the studied low-luminosity AGN (log L$_{X}<$ 42.5) shows evidence for significantly extended soft X-ray emission, which represents a possible contribution of the host galaxy to the X-ray luminosity. These cases are printed bold-face
in Tab. \ref{tab:RBS_sa}. The amount of soft extended emission for the 
low-luminosity RBS sample of AGN ranges from 10\% to 55\%. If the fraction of
extended emission is indeed a contribution of the surrounding galaxy this will
influence the faint end of the local XLF of AGN and the bright end of the XLF of local galaxies (cf. Fig. \ref{fig:XLF}).

In contrast, eight out of thirteen galaxies exhibit a clear spatially resolved source extension (between 75\% and 100\%) in the HRI band. The contribution of a possible hidden AGN to the total X-ray luminosity is very limited. As already mentioned, only a small number of galaxies are only slightly extended in soft X-rays (cf Fig. \ref{fig:rad_gal}). 

\begin{table*}[htp]
      \caption[]{The RBS low-luminosity sample of AGN (ID$=$1) and galaxies 
      (ID$=$2). CR and CH are the total and hard band (0.5$-$2.0 keV)
      ROSAT PSPC countrates, the X-ray luminosity L$_{X}$, the redshift and the
      fraction of extended X-ray emission. A horizontal bar indicates cases without sufficient HRI data. Bold numbers mark AGN with a significant extended X-ray emission.
      }
\begin{center}
\begin{tabular}{l|l|c|c|c|c|c|c}
\hline
 Name &  Type & ID & CR & CH & L$_{X}$ &  z & Fraction of extended\\
      &    &   & [cts/s] & [cts/s] & [ergs/s] &     & X-ray emission[\%]\\
\hline
Holmberg II &   Gal/HII   &2 &0.223 & 0.207&39.44&0.0005  & 27$\pm$5 \\
M 81        &   Sy1.8     &1 &0.998 & 0.933&39.66&0.0003  & {\bf 12$\pm$3} \\
M 94        &   Gal       &2 &0.255 & 0.125&39.80&0.0010  & 79$\pm$3 \\
NGC 0253    &   Gal/Starburst&2 &0.298 & 0.256&39.91&0.0008  & 92$\pm$1 \\
NGC 5905    &   Sy1       &1 &0.310 & 0.002&40.00&0.0113  &   $-$    \\
M 33        &   Gal       &2 &0.556 & 0.539&40.04&0.0006  & 33$\pm$3 \\
M82         &   Gal/Starburst  &2 &0.897 & 0.897&40.38&0.0007  & 97$\pm$4 \\
NGC 1313    &   Gal       &2 &0.227 & 0.227&40.50&0.0016  & 18$\pm$5\\
M 83        &   Gal       &2 &0.261 & 0.223&40.54&0.0017  & 90$\pm$7 \\
NGC 4151    &   Sy1       &1 &0.224 & 0.139&40.89&0.0033  & {\bf 32$\pm$2} \\
NGC 5033    &   Sy1.9     &1 &0.338 & 0.193&40.93&0.0030  & 13$\pm$8 \\
NGC 4203    &   Gal       &2 &0.521 & 0.326&41.32&0.0036  & 11$\pm$5 \\
NGC 1566    &   Sy1       &1 &0.425 & 0.185&41.37&0.0050  & {\bf 21$\pm$5} \\
NGC 3998    &   Sy1/Liner &1 &0.629 & 0.487&41.48&0.0035  &  6$\pm$4 \\
NGC 4051    &   Sy1       &1 &3.918 & 1.077&41.49&0.0024  & {\bf 10$\pm$2} \\
NGC 1404    &   Gal       &2 &0.305 & 0.201&41.63&0.0065  & 87$\pm$7 \\
RXS J1447+1145&   Gal       &2 &0.261 & 0.010&41.68&0.0300  & $-$    \\
NGC 4636    &   Gal       &2 &0.936 & 0.800&41.75&0.0037  &100$\pm$27\\
NGC 3147    &   Sy2       &1 &0.205 & 0.130&41.79&0.0094  & {\bf 25$\pm$6} \\
WPVS 07     &   Sy1/NL    &1 &0.959 & 0.014&41.80&0.0288  &  $-$   \\
NGC 1068    &   Sy2       &1 &1.785 & 0.866&41.82&0.0038  & {\bf 55$\pm$8} \\
UGC 06728   &   Sy1/NL    &1 &0.375 & 0.336&41.82&0.0060  &  7$\pm$6 \\
ESO 568- 21 &   Sy1/NL    &1 &0.241 & 0.117&41.92&0.0114  & 17$\pm$14\\
NGC 5846    &   Gal/E     &2 &0.467 & 0.460&41.97&0.0061  & 98$\pm$5 \\
NGC 5813    &   Gal/E     &2 &0.545 & 0.520&42.06&0.0064  & 97$\pm$11\\
MRK 1126    &   Sy1.5     &1 &0.350 & 0.199&42.08&0.0106  &  $-$     \\
MRK 1502    &   Sy1       &1 &0.818 & 0.609&42.09&0.0060  &  5$\pm$5 \\
IRAS 22146-5955&AGN       &1 &0.798 & 0.132&42.21&0.0153  &  $-$     \\
RXS J0439-5311 &   Sy1/NL    &1 &0.759 & 0.061&42.26&0.0243  & $-$   \\
MRK 202     &   Gal/Starburst &2 &0.213 & 0.085&42.28&0.0210  & 6$\pm$4 \\
ESO 548-81  &   AGN       &1 &0.258 & 0.172&42.28&0.0145  & $-$   \\
NGC 3035    &   Sy1       &1 &0.263 & 0.226&42.41&0.0144  &  7$\pm$7 \\
NGC 4619    &   Sy1       &1 &0.222 & 0.100&42.43&0.0231  & {\bf 17$\pm$7} \\
\hline
\end{tabular}
\end{center}
\label{tab:RBS_sa}
\end{table*} 

\section{Conclusions}
We have found evidence for a significant contribution of the host galaxy 
to the X-ray luminosity of AGN in a statistically complete RBS sample of 
low-luminosity AGN. Future spatially resolved hard X-ray spectroscopy
is neccessary in order to uncover the nature of this extended X-ray emission.

\begin{acknowledgements}
We made use of the ROSAT HRI public data archive at the MPE. The work has been supported in part by the DLR (former DARA GmbH) under grant
50~OR~9403~5.
\end{acknowledgements}

\end{document}